\documentclass[11pt]{article}
     
\usepackage{latexsym}
\usepackage{epsfig}

  \large
\begin{document}
\newcommand{\Dirac}{\rlap{\hspace{-.5mm} \slash} D}
\thispagestyle{empty} \hfill{}

\vspace{2cm}
\begin{center}
{\Large\bf QCD at small non-zero quark chemical potentials} \vspace{8mm}

J.B. Kogut and D. Toublan \vskip 0.2cm {\it Loomis Laboratory of
Physics, University of Illinois, Urbana-Champaign, IL 61801, USA}
\vskip 1.5cm

{\bf Abstract}
\end{center}
{\footnotesize We study the effects of small chemical potentials
associated with the three light quark flavors in QCD. We use a
low-energy effective field theory that solely relies on the symmetries
of the QCD partition function. We find
three different phases: a normal phase, a pion superfluid phase
and a kaon superfluid phase. The two superfluid phases are
separated by a first order phase transition, whereas the normal
phase and either of the superfluid phases are separated by a
second order phase transition. We compute the quark-antiquark
condensate, the pion condensate and the kaon condensate in each
phase, as well as the isospin density, the strangeness density, and the
mass spectrum.}

\vskip 0.5cm 
\noindent 
{\footnotesize {\it PACS}:11.30.Rd; 12.39.Fe; 12.38.Lg; 71.30.+h
\\  \noindent
{\it Keywords}: QCD phase diagram; Pion condensate; Kaon condensate; 
Low-energy effective field theory; Chiral Perturbation Theory} 

\vskip1.5cm
\noindent
\section{Introduction}

QCD at non-zero baryon density has recently been the subject of
many studies. Our present knowledge of nucleon rich matter
is limited to low densities and very high densities. In
the first regime, we can rely on our good phenomenological
understanding of nuclear interactions in the vacuum. In the second
regime, the phenomenon of color superconductivity and its
consequences are becoming better understood both qualitatively and
quantitatively \cite{CSC0,CSC1,CSC2,RWreview}. However we have only a poor
grasp of a sample of nuclear matter that is neither too dilute nor
too dense. Unfortunately this kind of matter is found in many interesting
physical systems, from supernova
explosions and neutron stars to heavy ion collisions.
In other words, our knowledge of the QCD phase diagram at
non-zero chemical potential is limited to extreme cases, and it
is clear that we need a better understanding of its characteristics.

This situation is very different from what we know about 
QCD at non-zero temperature and zero chemical potential, 
where numerous analytic and numerical studies have enabled
us to achieve a qualitative and a quantitative understanding 
\cite{TQCD}. Unfortunately, the current numerical
techniques are useless in the case of QCD with three colors and
fermions in the fundamental representation at non-zero 
baryon density. The complexity of the
measure in the partition function is the origin of the problem.
Special techniques have been developed to tackle this problem,
but their reach is presently far from QCD \cite{Meron}. 
There are however QCD-like theories at non-zero baryon
density that do not have this complexity problem. Therefore they can be studied
numerically with the usual techniques:
QCD with two colors and fundamental fermions, and QCD
with any number of colors and adjoint fermions. We now have a good
understanding of these QCD-like theories from analytic and
numerical studies \cite{KST,KSTVZ,SSS,Simon,Aloisio,Liu,SimonJohn}. 
QCD with three colors and two flavors of 
fundamental fermions at non-zero isospin density also has a real measure,
and can therefore also be studied with standard numerical algorithms 
\cite{SonMisha}.

In this article we will study the QCD phase diagram for small
chemical potentials associated to each quark flavor. 
At low enough energies, the single most important property
of QCD is that chiral symmetry is spontaneously broken in the 
vacuum. It implies the existence of light bosonic degrees
of freedom, and stringent constraints on the physical observables. 
This property has been extensively 
and successfully used to study QCD at low energy within an 
effective theory solely based on the symmetries of the QCD partition function,
Chiral Perturbation Theory \cite{W,GL}. 
Since the spontaneous breaking of chiral symmetry is still the most relevant 
property of QCD at low enough quark chemical potentials, Chiral Perturbation 
Theory should also enable us to explore the QCD phase diagram in this domain.
This approach has been successfully used for QCD-like theories at non-zero
baryon chemical potential and for QCD at non-zero isospin chemical potential
\cite{KST,KSTVZ,SSS,SonMisha}. 

We will restrict ourselves to the three light quark flavors, for which this 
effective theory approach can be used. There is however
an intrinsic limitation to this approach. Since only the octet of Goldstone 
bosons are kept as the relevant degrees of freedom, high chemical 
potentials where other excitations become relevant cannot be reached. 
A rapid glance at the hadron spectrum, and in particular the proton, the rho 
and the omega, shows that we will be able to explore the domain restricted by
$|\mu_u|<300 {\rm MeV}$,
$|\mu_d|<300 {\rm MeV}$, and $|\mu_s|<550 {\rm MeV}$.

Our study extends the analysis of
QCD at low isospin chemical potential that included the up and
down quark flavors only \cite{SonMisha}. 
In this analysis, it was found that the pions form a condensate
at high enough isospin chemical potential. Similarly, we
will find that for high enough strange chemical potential, 
it is energetically favorable for the kaons to form a condensate.
A non-trivial phase diagram emerges; there are three distinct phases:  
a normal phase, a pion condensation phase and a kaon condensation phase. 
The superfluid phases are separated from each other by a first order
phase transition, whereas either of them is separated from the
normal phase by a second order phase transition. 

The pion and kaon condensates we will find in our study are very different from
the pion and kaon condensates that emerge in a nucleon rich environment 
\cite{PiKCondNucl}. In this latter case, the condensation is driven by the 
attractive meson-nucleon interactions. In our case the baryon density is zero, 
and the pion and kaon condensates emerge out of the vacuum
as soon as it becomes energetically favorable for the system to form
such condensates in the ground state, just because of the charge carried by these modes.

We proceed as follow.
First we shortly derive the effective Lagrangian in Section~2, and then determine
the ground state of the effective theory in Section~3. Section~4
is devoted to the computation of the various condensates relevant
to the three different phases as well as the isospin and
strangeness densities. The mass spectrum is computed in Section~5.
Concluding remarks and discussion are presented in the Section~6.

\section{Low-energy Effective Lagrangian}

In order to construct the low-energy effective Lagrangian, we need
to study the symmetries of QCD. We work in
Minkowski space. The fermionic part of the QCD Lagrangian at
non-zero quark chemical potentials is given by
\begin{equation}
{\cal L}_{\rm QCD}=\bar \psi \Dirac \psi+\bar \psi B_\nu
\gamma_\nu \psi - \bar \psi {\cal M} \psi,
\end{equation}
where
\begin{eqnarray}
\psi=\left ( \begin{array}{c} u \\ d \\ s \end{array} \right ), \hspace{1cm}
{\cal M}={\rm diag}(m_u,m_d,m_s), 
\end{eqnarray}
\begin{eqnarray}
{\rm and} \; \; B_\nu=(-B,\vec{0}), \; \; B={\rm diag}(\mu_u,\mu_d,\mu_s)={\rm
diag} (\frac13 \mu_B+\frac12 \mu_I,\frac13 \mu_B-\frac12 \mu_I, \frac13
\mu_B-\mu_S).
\end{eqnarray}
One can either use chemical potentials associated with the quark
flavors, or chemical potentials associated with baryon number,
$\mu_B$, isospin, $\mu_I$, and strangeness, $\mu_S$. In this
study, we will take $m_u=m_d \equiv m$. When all quark masses and
chemical potentials vanish, the QCD Lagrangian is invariant under
$SU_L(3) \times SU_R(3) \times U_V(1)$. At zero chemical
potential, this symmetry is spontaneously broken to $SU_V(3)
\times U_V(1)$. There are therefore eight Goldstone bosons in the
spectrum: $\pi^0$, $\pi^+$, $\pi^-$, $\eta^0$, $K^+$, $K^-$,
$K^0$, and $\bar K^0$. They are the relevant degrees of freedom at
low energy.

The effective Lagrangian of QCD at low energies and finite quark
chemical potentials in Minkowski space is given by
\begin{eqnarray}
\label{Leff}
{\cal L}_{\rm eff}=\frac{F^2}{4} {\rm Tr} \nabla_\nu \Sigma
\nabla_\nu \Sigma^\dagger+\frac12 G {\rm Tr} {\cal M}
(\Sigma+\Sigma^\dagger),
\end{eqnarray}
where $\Sigma \in SU(3)$, and $\nabla_\nu \Sigma=\partial_\nu
\Sigma - i [B_\nu,\Sigma]$. This is the usual Chiral Perturbation
Theory Lagrangian at lowest order \cite{W,GL}. It contains only
the octet of Goldstone bosons due to spontaneous chiral symmetry breaking.
The chemical potential has been introduced in the same way as a regular vector source
using the usual flavor gauge symmetry \cite{GL,KST}. Under a chiral
rotation, the quark fields transform as $(\psi_L,\psi_R)
\rightarrow (V_L \psi_L, V_R \psi_R)$, with $V_{L,R} \in
SU(3)_{L,R}$, whereas $\Sigma \rightarrow V_R \Sigma V_L^\dagger$.
For zero chemical potentials $\Sigma$ is
invariant under vector transformations and contains the
octet of Goldstone bosons due to spontaneous chiral symmetry
breaking:
\begin{eqnarray}
\Sigma=U \bar \Sigma U, \; {\rm with} \; U={\rm exp}(i \Phi/\sqrt{2} F),
\label{Sig}
\end{eqnarray}
and
\begin{eqnarray}
\Phi=\left( \begin{array}{ccc}
\frac{\pi^0}{\sqrt{2}}+\frac{\eta^0}{\sqrt{6}} & \pi^+ & K^+\\
\pi^- & -\frac{\pi^0}{\sqrt{2}}+\frac{\eta^0}{\sqrt{6}} & K^0 \\
K^- & \bar K^0 & -\frac{2 \eta^0}{\sqrt{6}}
 \end{array} \right), \label{Goctet}
\end{eqnarray}
with $\bar \Sigma={\rm diag}(1,1,1)$ for
$\mu_B=\mu_I=\mu_S=0$. The effective Lagrangian is invariant under local
chiral transformations. At zero chemical potentials, the masses of the
Goldstone bosons are given by the Gell-Mann--Oakes--Renner relation:
\begin{eqnarray}
\left.
\begin{array}{lll}
  m_\pi^2&=&2 G m/F^2  \\
  m_K^2&=&G (m+m_s)/F^2 \\
  m_{\eta^0}^2&=&2 G (m+2 m_s)/3 F^2=(4 m_K^2-m_\pi^2)/3.
\end{array}
\right.
\end{eqnarray}

The first remarkable property of the effective Lagrangian
(\ref{Leff}) is that in the $(\mu_B,\mu_I,\mu_S)$-basis, $\mu_B$
completely drops out of ${\cal L}_{\rm eff}$. This is just due to
the fact that the effective Lagrangian contains only the octet of
Goldstone bosons due to spontaneous chiral symmetry breaking. The
use of the effective Lagrangian is therefore limited to
$|\mu_B|<940 {\rm MeV}$, $|\mu_I|<770 {\rm MeV}$, and $|\mu_S|<550
{\rm MeV}$. These estimated bounds come from the masses of the
proton, the rho and the omega, respectively. This means that the effective
Lagrangian is only valid in the domain $|\mu_u|<300 {\rm MeV}$, 
$|\mu_d|<300 {\rm MeV}$, and $|\mu_s|<550 {\rm MeV}$. Therefore in this 
work, the baryon chemical potential will never appear explicitly, and we will 
work with $\mu_I$ and $\mu_S$. Recall however that
all our results are valid for $|\mu_B|<940 {\rm MeV}$, and that
Chiral Perturbation Theory, in its domain of validity, allows us to
determine the phase diagram in the {\it three} quark chemical potentials.
For definiteness, we will restrict
ourselves to $\mu_I,\mu_S>0$, the other quadrants in the
$(\mu_I,\mu_S)$-plane are easily derived from our results.

\section{Ground state}
The ground state of the effective theory is determined by the
maximum of the static part of the effective Lagrangian. Therefore
$\bar \Sigma$, the maximum of
\begin{eqnarray}
{\cal L}_{\rm stat}&=&-\frac{F^2}{4} {\rm Tr} [B,\bar \Sigma]
[B,\bar \Sigma^\dagger]+\frac12 G {\rm Tr} {\cal M}
(\bar \Sigma+\bar \Sigma^\dagger), \label{Lstat}
\end{eqnarray}
determines the ground state of QCD at non-zero quark chemical
potentials.

We will use the following Ansatz for the saddle point:
\begin{eqnarray}
\bar \Sigma=\left( \begin{array}{ccc} 1 & 0 & 0 \\
                                 0 & \cos \beta & -\sin \beta \\
                                 0 & \sin \beta & \cos \beta
                   \end{array} \right) \;
            \left( \begin{array}{ccc} \cos \alpha & \sin \alpha  & 0 \\
                                 -\sin \alpha & \cos \alpha & 0 \\
                                 0 & 0 & 1
                   \end{array} \right) \;
            \left( \begin{array}{ccc} 1 & 0 & 0 \\
                                 0 & \cos \beta & \sin \beta \\
                                 0 & -\sin \beta & \cos \beta
                   \end{array} \right),
\label{SadP}
\end{eqnarray}
with $\alpha,\beta \in (0,\pi/2)$. Notice that $\bar
\Sigma|_{\alpha=0}={\rm diag}(1,1,1)$ corresponds to the
ground state at $\mu_I=\mu_S=0$. Furthermore in the massless
case, 
\begin{eqnarray}
  \label{piOrient}
  \bar \Sigma|_{(\alpha=\frac \pi2,\beta=0)}=
      \left( \begin{array}{ccc}  0 & 1 & 0 \\
                                          -1 & 0 & 0 \\
                                            0 & 0 & 1
              \end{array} \right)
\end{eqnarray}
is found to be the maximum when $\mu_I>2 \mu_S$, and 
\begin{eqnarray}
  \label{KOrient}
  \bar
\Sigma|_{(\alpha=\frac \pi2,\beta=\frac \pi2)}=
      \left( \begin{array}{ccc}  0 & 0 & 1 \\
                                            0 & 1 & 0 \\
                                          -1 & 0 & 0
              \end{array} \right)
\end{eqnarray}
is found to be the maximum when
$\mu_I<2 \mu_S$. By 
noticing that the condensate is basically given by
$\bar \psi_R \bar \Sigma \psi_L$, it is easy
to see that (\ref{piOrient}) corresponds to a pion
condensation phase, and that (\ref{KOrient}) corresponds
to a kaon condensation phase.
Our Ansatz corresponds to an axial rotation to go from
the normal phase to either of the superfluid phases, and a vector
rotation to go directly from one superfluid phase to the other. We
will show that this Ansatz indeed corresponds to a local maximum
of ${\cal L}_{\rm stat}$. We will assume that it is the global
maximum.

After our Ansatz (\ref{SadP}) is introduced into (\ref{Lstat}),
the static part
of the effective Lagrangian becomes quite complicated
\begin{eqnarray}
{\cal L}_{\rm stat}&=&\frac{1}{32 F^2} \Big\{ \Big(21 \mu_I^2+12
\mu_I \mu_S+20 \mu_S^2+ (19 \mu_I^2+20 \mu_I \mu_S+12 \mu_S^2)
\cos \alpha \Big) \sin^2(\alpha/2) \nonumber \\
&& \hspace{1cm}-2 (\mu_I-2 \mu_S)^2 \sin^4(\alpha/2) \cos (4
\beta) + 2
(\mu_I-2 \mu_S) (3 \mu_I+2 \mu_S) \sin^2 \alpha \Big\} \\
&& +G \Big\{ m \cos \alpha (1+ \cos^2 \beta)  + m_s \cos^2 \beta +
( m+m_s \cos \alpha ) \sin^2 \beta \Big\}. \nonumber
\end{eqnarray}
We first maximize ${\cal L}_{\rm stat}$ with respect to $\beta$. Remember
that we limit ourselves to the first quadrant in the $(\mu_I,\mu_S)$-plane:
$\mu_I,\mu_S>0$.
We find that the maximum, for $\alpha\neq0$ and $\beta \in (0,\pi/2)$, is
either at $\beta=0$ or at $\beta=\pi/2$. Therefore there are two
different phases according to the value of $\beta$. We look for the
maximum of ${\cal L}_{\rm stat}$ separately in each case.
\renewcommand{\labelenumi}{\roman{enumi})}
\begin{enumerate}
\item \underline{$\beta=0$}: The static part of the effective Lagrangian, its
maximum, and the saddle point are given by
\begin{eqnarray}
{\cal L}_{\rm stat}&=&2 G m \cos \alpha +G m_s +
\frac{F^2}2 \mu_I^2 \sin^2 \alpha, \\
  {\rm with}&& \left\{
  \begin{array}{lll}
    \cos \alpha=1 & {\rm for} & \mu_I<m_\pi \\
    \cos \alpha=(m_\pi/\mu_I)^2 & {\rm for} & \mu_I>m_\pi,
  \end{array}
  \right.\\
  {\rm and} && \bar \Sigma=\left( \begin{array}{ccc}
                            \cos \alpha & \sin \alpha & 0 \\
                            -\sin \alpha & \cos \alpha & 0\\
                            0 & 0 & 1
                            \end{array} \right).
\end{eqnarray}

\item \underline{$\beta=\pi/2$}: The static part of the effective
Lagrangian, its maximum, and the saddle point are given by
\begin{eqnarray}
{\cal L}_{\rm stat}&=&G m +G(m+m_s) \cos \alpha +\frac{F^2}{2}
(\frac12 \mu_I+\mu_S)^2 \sin^2 \alpha,\\
  {\rm with} && \left\{
  \begin{array}{lll}
    \cos \alpha=1 & {\rm for} & \frac12 \mu_I+\mu_S<m_K \\
    \cos \alpha=(m_K/(\frac12 \mu_I+\mu_S))^2 & {\rm for}
    & \frac12 \mu_I+\mu_S>m_K,
  \end{array}
  \right.\\
  {\rm and} &&\bar \Sigma=\left( \begin{array}{ccc}
                            \cos \alpha & 0 & \sin \alpha  \\
                            0 & 1 & 0\\
                            -\sin \alpha & 0 & \cos \alpha
                            \end{array} \right).
\end{eqnarray}
\end{enumerate}

Finally, in order to find the maximum of the static part of
the effective Lagrangian in the whole quadrant $\mu_I,\mu_S>0$,
we have to compare the value of ${\cal L}_{\rm stat}$
for $\alpha=0$, for $\beta=0$ and $\cos \alpha=(m_\pi/\mu_I)^2$,
and for  $\beta=\pi/2$ and $\cos \alpha=(m_K/(\frac12 \mu_I+\mu_S))^2$.
The orientation of the condensate for a given pair $(\mu_I,\mu_S)$
is determined by the maximum of ${\cal L}_{\rm stat}$ at this point.
We find three different phases:
\renewcommand{\labelenumi}{\arabic{enumi}.}
\begin{enumerate}
\item {\bf Normal phase}: $\mu_I<m_\pi$ and $\mu_S<m_K-\frac12 \mu_I$
\begin{eqnarray}
    \cos \alpha=1, \; {\rm any} \;  \beta \in (0,\pi/2) \label{normal}
\end{eqnarray}
\item {\bf Pion condensation phase}: $\mu_I>m_\pi$ and $\mu_S<
\left(-m_\pi^2+\sqrt{(m_\pi^2-\mu_I^2)^2+4 m_K^2 \mu_I^2 } \right)/2 \mu_I$
\begin{eqnarray}
    \cos \alpha=\left(\frac{m_\pi}{\mu_I}\right)^2, \;  \beta=0 \label{pion}
\end{eqnarray}
\item {\bf Kaon condensation phase}: $\mu_S>m_K-\frac12 \mu_I$ and $\mu_S>
\left(-m_\pi^2+\sqrt{(m_\pi^2-\mu_I^2)^2+4 m_K^2 \mu_I^2 }\right)/2 \mu_I$
\begin{eqnarray}
    \cos \alpha=\left( \frac{m_K}{\frac12 \mu_I+\mu_S}\right)^2,
\; \beta=\pi/2. \label{kaon}
\end{eqnarray}
\end{enumerate}

The curve $\mu_S=\left(-m_\pi^2+\sqrt{(m_\pi^2-\mu_I^2)^2+4 m_K^2
\mu_I^2 } \right)/2 \mu_I$ is a first order phase transition
curve, because the pion and kaon condensates compete on this
curve. The phase diagram in the first quadrant and its obvious
extension to the whole $(\mu_I,\mu_S)$-plane are given in Fig.~1
for $m_\pi=140 {\rm MeV}$, and $m_K=500 {\rm MeV}$. They are very similar to
the phase diagram of QCD-like theories at finite baryon and isospin densities \cite{SSS}.
\begin{center}
\begin{figure}[ht!]
\vspace{0.5cm}
\hspace*{1cm}
\epsfig{file=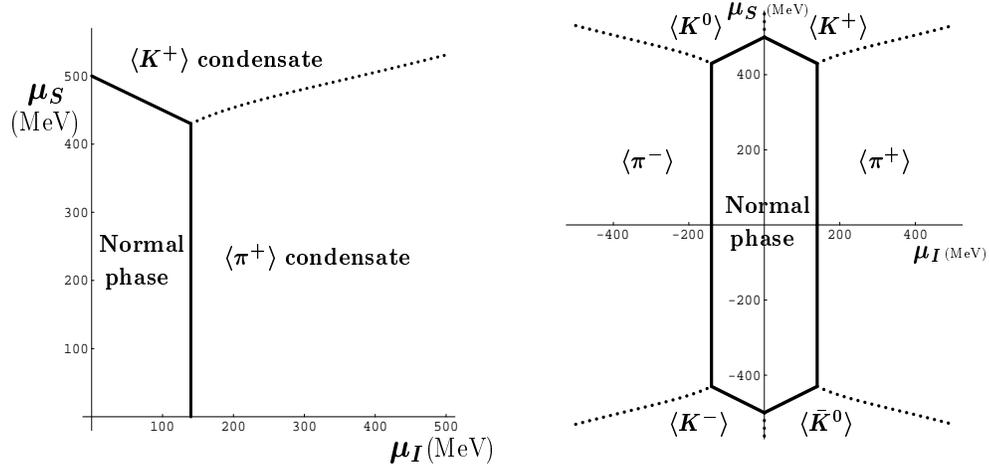, width=14.6cm, height=9.5cm}
\vspace{-3cm}
\caption[]{\small Phase diagram in the $(\mu_I,\mu_S)$-plane. The solid
lines are second order phase transitions, whereas the dotted curves are
first order phase transitions. The intersection of the three phase
transition curves on the left figure is at
$(\mu_I,\mu_S)=(m_\pi,m_K-m_\pi/2)$.}
\end{figure}
\end{center}

\vspace{-1cm}
\section{Condensates and charge densities}
We will show in the next section that $\bar \Sigma$ given by
(\ref{SadP},\ref{normal},\ref{pion},\ref{kaon}) is indeed a local
maximum of the static part of the effective Lagrangian. We will
assume that it is the global maximum. But first we investigate the
properties of the ground state described by
(\ref{SadP},\ref{normal},\ref{pion},\ref{kaon}) in the three
phases by looking at different observables. The quark-antiquark
condensates, $\langle \bar q q \rangle$ for $q=u,d,s$,  the pion
condensate,  $\langle \bar d \gamma_5 u + {\rm h.c.}\rangle$, the
kaon condensate, $\langle \bar s \gamma_5 u + {\rm h.c.} \rangle$ 
are obtained by inserting the appropriate sources into the effective
Lagrangian \cite{KSTVZ}, or more directly by noticing that 
the vacuum expectation value of an operator bilinear in the quark fields 
is basically given by $\langle \bar{\psi}_R \bar \Sigma \psi_L \rangle$
(\ref{piOrient},\ref{KOrient}).
The isospin density, $n_I=\partial {\cal L}_{\rm eff}/\partial
\mu_I$, and  the strangeness density, $n_S=\partial {\cal L}_{\rm
eff}/\partial \mu_S$ are easily obtained from the static part of the
effective Lagrangian.

\begin{enumerate}
\item {\bf Normal phase}, $\mu_I<m_\pi$ and $\mu_S<m_K-\frac12 \mu_I$:
\begin{eqnarray} \left\{
\begin{array}{l}
\langle \bar u u \rangle=\langle \bar d d \rangle=\langle \bar s s \rangle=G
 \equiv \langle \bar q q \rangle_0 \\
\langle \bar d \gamma_5 u + {\rm h.c.} \rangle=
\langle \bar s \gamma_5 u + {\rm h.c.} \rangle=0 \\
n_I=n_S=0
\end{array} \right.
\end{eqnarray}

\item {\bf Pion condensation phase}, $\mu_I>m_\pi$ and $\mu_S<\left(-m_\pi^2+
\sqrt{(m_\pi^2-\mu_I^2)^2+4 m_K^2 \mu_I^2 }\right)/2 \mu_I$:
\begin{eqnarray} \left\{
\begin{array}{l}
\langle \bar u u \rangle=\langle \bar d d \rangle=G \cos \alpha =
\langle \bar q q \rangle_0 m_\pi^2/\mu_I^2 \\
\langle \bar s s \rangle=G=\langle \bar q q \rangle_0 \\
\langle \bar d \gamma_5 u + {\rm h.c.} \rangle
=G \sin \alpha=\langle \bar q q \rangle_0
\sqrt{1-m_\pi^4/\mu_I^4} \\
\langle \bar s \gamma_5 u + {\rm h.c.} \rangle=0 \\
n_I=F^2 \mu_I (1-m_\pi^4/\mu_I^4)\\
n_S=0
\end{array} \right.
\end{eqnarray}

\item {\bf Kaon condensation phase}, $\mu_S>m_K-\mu_I/2$ and $\mu_S>\left(-m_\pi^2+
\sqrt{(m_\pi^2-\mu_I^2)^2+4 m_K^2 \mu_I^2 }\right)/2 \mu_I$:
\begin{eqnarray} \left\{
\begin{array}{l}
\langle \bar u u \rangle=\langle \bar s s \rangle=G \cos \alpha =
\langle \bar q q \rangle_0 m_K^2/(\frac12 \mu_I^2+\mu_S)^2 \\
\langle \bar d d \rangle=G=\langle \bar q q \rangle_0 \\
\langle \bar d \gamma_5 u + {\rm h.c.} \rangle=0 \\
\langle \bar s \gamma_5 u + {\rm h.c.} \rangle
=G \sin \alpha=\langle \bar q q \rangle_0
\sqrt{1-m_K^4/(\frac12 \mu_I+\mu_S)^4} \\
n_I=\frac12 F^2 (\frac12 \mu_I+\mu_S) \Big(1-m_K^4/(\frac12 \mu_I+\mu_S)^4
\Big)\\
n_S=F^2 (\frac12 \mu_I+\mu_S) \Big(1-m_K^4/(\frac12 \mu_I+\mu_S)^4 \Big).
\end{array} \right.
\end{eqnarray}
\end{enumerate}
\begin{center}
\begin{figure}[ht!]
\vspace{0.5cm} \hspace*{1cm} \epsfig{file=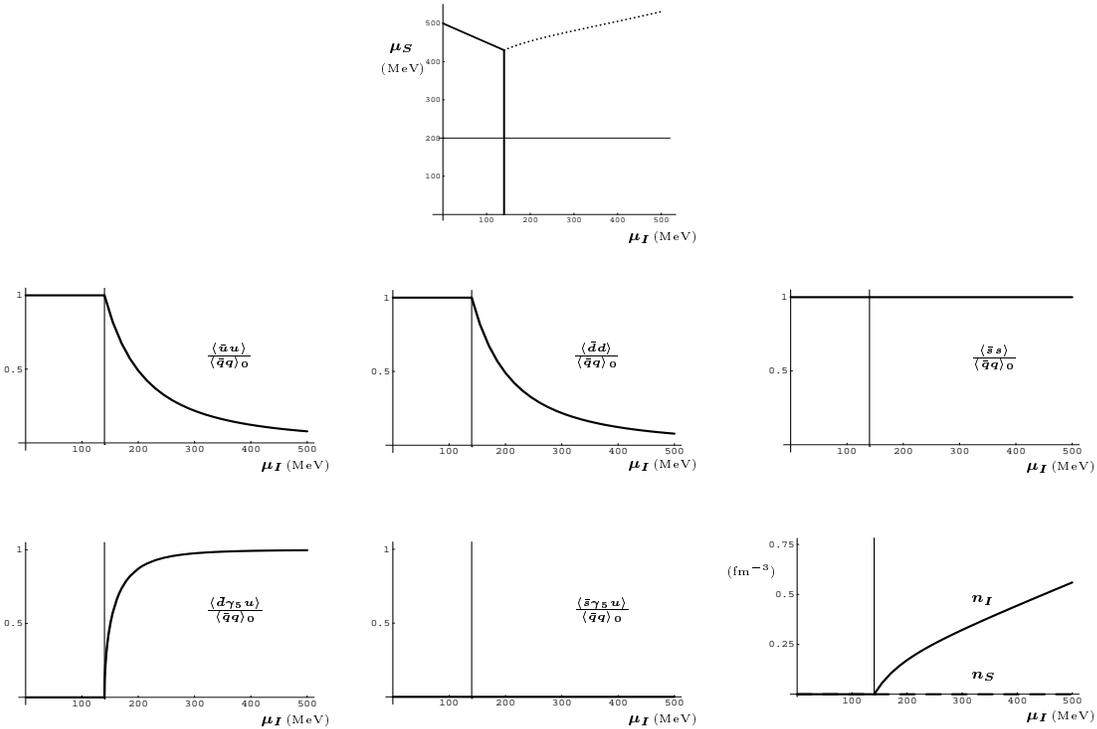,
width=14.3cm, height=12cm} \vspace{-3cm} \caption[]{\small The
figure on top is the phase diagram in the first quadrant of the
$(\mu_I,\mu_S)$-plane. The light solid horizontal line corresponds to
$\mu_S=200$MeV, the value we use throughout the other figures
depicted underneath. In these last six figures, the critical value
of $\mu_I$ at the second order phase transition is depicted by a
solid vertical line. The different condensates are shown in the
next five figures. The isospin density (solid curve) and the
strangeness density (dashed curve) are shown together in the last
figure.}
\end{figure}
\end{center}
The isospin and strangeness densities in the superfluid
phases just above the second order phase transition curves
can be reproduced within a semi-classical analysis of a dilute
Bose gas in the Bose-Einstein condensation phase \cite{KSTVZ}. In the 
kaon superfluid phase, the
strangeness density is naturally found to be twice bigger than the isospin
density, since a kaon carries twice more strangeness than isospin.

The different condensates and charge densities as a function of
$\mu_I$ and for a constant $\mu_S$ are depicted in Fig.~2, for
$\mu_S=200$MeV, and in Fig.~3, for $\mu_S=460$MeV. In the first
case we encounter only two phases as $\mu_I$ changes, whereas in
the second case we encounter the three different phases found
above as $\mu_I$ changes. As expected, the condensates and charge
densities are continuous across the second order phase transition
and discontinuous across the first order one. The different phases and the nature
of the different phase transitions are
clearly distinguished by the observables we study.

\begin{center}
\begin{figure}[ht!]
\vspace{0.5cm} \hspace*{1cm} \epsfig{file=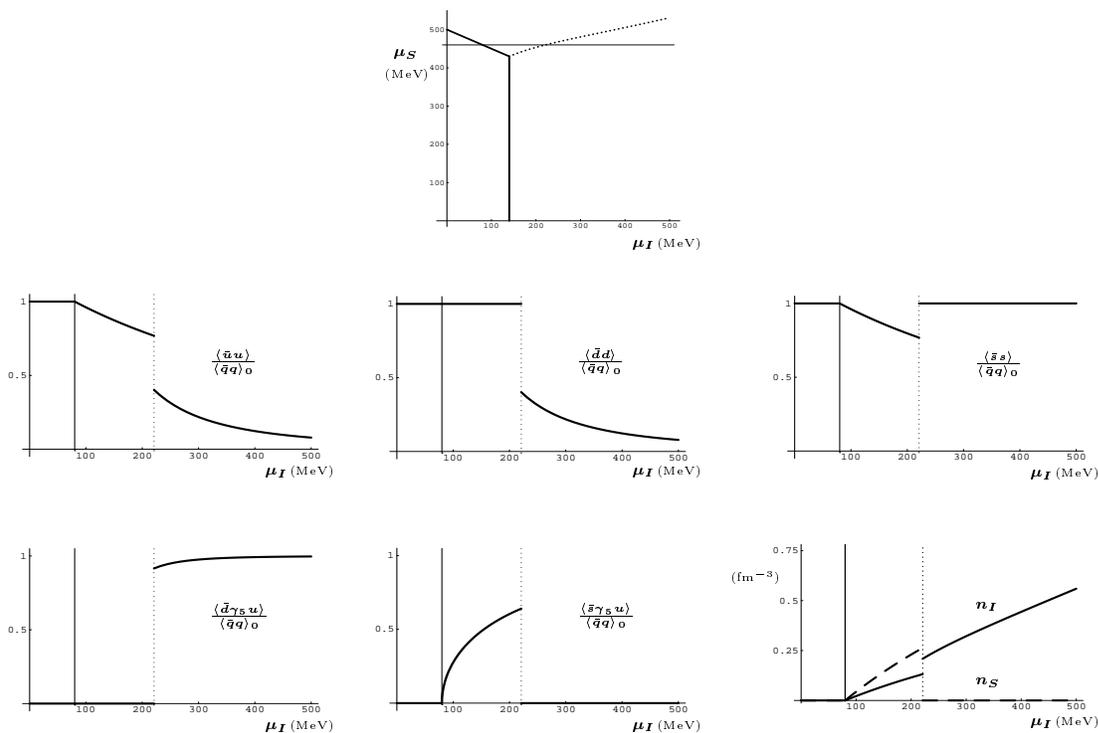,
width=14.3cm, height=12cm} \vspace{-3cm} \caption[]{\small The
figure on top is the phase diagram in the first quadrant of the
$(\mu_I,\mu_S)$-plane. The light solid horizontal line corresponds to
$\mu_S=460$MeV, the value we use throughout the other figures
depicted underneath. In these last six figures, the critical value
of $\mu_I$ at the second order phase transition is depicted by a
solid vertical line, and the critical value of $\mu_I$ at the
first order phase transition is depicted by a dotted vertical
line. The different condensates are shown in the next five
figures. The isospin density (solid curve) and the strangeness
density (dashed curve) are shown together in the last figure. The
difference between the second and first order phase transitions are very
clearly seen in these observables.} \end{figure}
\end{center}

\section{Curvatures and mass spectrum}
In this section we will prove that $\bar \Sigma$ given by
(\ref{SadP},\ref{normal},\ref{pion},\ref{kaon})
is indeed a local maximum of the static part of the effective Lagrangian.
We compute the masses of the different excitations in
the three phases. In order to obtain these masses, we have to expand the
whole effective Lagrangian up to second order in the Goldstone fields.
The general form of the effective Lagrangian reads
\begin{eqnarray}
\label{Lexpl}
  {\cal L}_{\rm eff}&=&-\frac{F^2}{4} {\rm Tr} [B,\bar \Sigma]
[B,\bar \Sigma^\dagger]+\frac12 G {\rm Tr} {\cal M}
(\bar \Sigma+\bar \Sigma^\dagger) \nonumber \\
&&+\frac{F^2}{4} {\rm Tr} \partial_\nu \Sigma
\partial_\nu \Sigma^\dagger
+ i \frac{F^2}{2} {\rm Tr} B [\Sigma^\dagger, \partial_0 \Sigma].
\end{eqnarray}
The fourth term in (\ref{Lexpl}) may involve a term quadratic in the
Goldstone fields (\ref{Goctet}). In the superfluid phases, the
original Goldstone fields are mixed \cite{KSTVZ}. Since the usual
Goldstone manifold changes as the normal phase is left, the
original octet of Goldstone bosons is not an appropriate set of
coordinates for the new Goldstone manifold. We will denote these
new appropriate fields by a ``tilde''. We choose them so that the
kinetic term in the Lagrangian is canonical. Furthermore, the
linear derivative term in (\ref{Lexpl}) mixes these new fields.
Therefore the mass spectrum is not directly given by the
curvatures of the static part of the Lagrangian at the maximum. 
When $n$-fields are mixed, the inverse propagator that appears
in the effective Lagrangian is a $n \times n$ matrix, and the masses are
given by the zeros of its determinant in the $p_0^2$-plane. A secular
equation has therefore to be solved. It is a $n$-th order equation in
$p_0^2$ when $n$ fields are mixed by the linear derivative term in
(\ref{Lexpl}). These equations can be explicitly solved, but the
explicit solutions are cumbersome and rather useless expressions.
Therefore we will only give which fields are mixed by the linear
derivative term in (\ref{Lexpl}), and solve the secular equation
explicitly in the simple cases only. The labels of the different mixed states will be
given according to the usual Goldstone mode
they correspond to at the second order phase transition
with the normal phase. We study each phase separately, and summarize
our results for the mass spectrum in a few figures.

\subsection{Normal phase}
For $\mu_I<m_\pi$ and $\mu_S<m_K-\frac12 \mu_I$, with the Goldstone
fields defined in (\ref{Goctet}), we find that there is no mixing
induced by the linear derivative term
in the quadratic part of the effective Lagrangian. The various masses read
\begin{eqnarray}
  \begin{array}{lll}
    m_{\pi^0}&=&m_\pi \\
    m_{\eta^0}&=&m_{\eta^0}=
      \sqrt{(4 m_K^2-m_\pi^2)/3} \\
    m_{\pi^+}&=&m_\pi -\mu_I\\
    m_{\pi^-}&=&m_\pi +\mu_I \\
    m_{K^+}&=&m_K-\frac12 \mu_I-\mu_S  \\
    m_{K^-}&=&m_K+\frac12 \mu_I+\mu_S \\
    m_{K^0}&=&m_K+\frac12 \mu_I-\mu_S  \\
    m_{\bar K^0}&=&m_K-\frac12 \mu_I+\mu_S.
  \end{array}
\end{eqnarray}
The effect of the chemical potentials is just to shift the mass of a state
by its charges times the chemical potentials: $E \rightarrow E-I \mu_I-S
\mu_S$, where $I$ and $S$ are the isospin and strangeness of
that state.

\subsection{Pion condensation phase}
For $\mu_I>m_\pi$ and $\mu_S<\left(-m_\pi^2+\sqrt{(m_\pi^2-\mu_I^2)^2+4 m_K^2
\mu_I^2 }\right)/2 \mu_I$, we find that there is mixing between some
of the fields defined in (\ref{Goctet}), namely $(\pi^0, \eta^0, \pi^+, \pi^-)$, and
$(K^+,  K^-, K^0, \bar K^0)$.
We find that the different masses
are given by
\begin{eqnarray}
  \begin{array}{lll}
    m^2_{\tilde \pi^0}&=&\mu_I^2 \\
    m_{\tilde \eta^0}^2&=&
      \frac1{6 \mu_I^2} \Big(  4 m_K^2 \mu_I^2-2 m_\pi^2 \mu_I^2
          +6 m_\pi^4+7 \mu_I^4
        + \Big[48 m_\pi^6 (\mu_I^2-m^2_\pi) \\
          &&  +\mu_I^4 (4 m_K^2+\mu_I^2)^2
            -4 m_\pi^4 \mu_I^2 (24 m_K^2-43 \mu_I^2)
            -4 m_\pi^2 \mu_I^4 (4 m_K^2+\mu_I^2) \Big]^{1/2}   \Big) \\
    m_{\tilde \pi^+}^2&=&0  \\
    m_{\tilde \pi^-}^2&=&
      \frac1{6 \mu_I^2} \Big(  4 m_K^2 \mu_I^2-2 m_\pi^2 \mu_I^2
          +6 m_\pi^4+7 \mu_I^4
        - \Big[48 m_\pi^6 (\mu_I^2-m^2_\pi) \\
          &&  +\mu_I^4 (4 m_K^2+\mu_I^2)^2
            -4 m_\pi^4 \mu_I^2 (24 m_K^2-43 \mu_I^2)
            -4 m_\pi^2 \mu_I^4 (4 m_K^2+\mu_I^2) \Big]^{1/2}   \Big),
\end{array}
\end{eqnarray}
and the masses of $\tilde K^+, \tilde K^-, \tilde K^0, \tilde {\bar K}^0$ are
given by the zeros on the $E$-plane of
\begin{eqnarray}
{\rm det} \left(
\begin{array}{cccc}
\frac12 E^2 + s_1 & E  s_2 & E s_3 & 0 \\
 E s_2^\dagger &
  \frac12 E^2 + s_1  & 0 &  E s_3 \\
E s_3^\dagger & 0 &
  \frac12 E^2 + s_1 & E s_4 \\
0 & E s_3^\dagger & E s_4^\dagger&
  \frac12 E^2 + s_1
  \end{array}
\right),
\end{eqnarray}
where
\begin{eqnarray}
  \begin{array}{lll}
    s_1&=&\frac12  ( - m_K^2 - \frac14 \mu_I^2 + \mu_S^2 +
          \mu_I( \frac12 \mu_I + \mu_S) \cos \alpha )\\
    s_2&=&  i  (\mu_S + \frac12 \mu_I \cos \alpha ) \\
    s_3&=&\frac12 \mu_I \sin \alpha \\
    s_4&=& i (\mu_S - \frac12 \mu_I \cos \alpha ),
  \end{array}
\end{eqnarray}
and $\cos \alpha=(m_\pi/\mu_I)^2$, as given in (\ref{pion}).

\subsection{Kaon condensation phase}
For $\mu_S>m_K-\mu_I/2$ and $\mu_S>\left(-m_\pi^2+\sqrt{(m_\pi^2-\mu_I^2)^2+4
m_K^2 \mu_I^2 }\right)/2 \mu_I$,we find that there is mixing between some
of the fields defined in (\ref{Goctet}), namely $(\pi^0, \eta^0,
K^+, K^-)$, and
$(\pi^+, \pi^-, K^0, \bar K^0)$.
The masses of $\tilde \pi^0, \tilde \eta^0,
\tilde K^+, \tilde K^-$ are given by the zeros in the $E$-plane of
\begin{eqnarray}
{\rm det} \left(
  \begin{array}{cccc}
\frac12 E^2 + t_{11} & t_{12} & 0 & 0 \\
 t_{12}^\dagger &
  \frac12 E^2 + t_{22}  & E t_{23} &  0 \\
 0 & E t_{23}^\dagger &
  \frac12 E^2 & E t_{34} \\
0 & 0 & E t_{34}^\dagger&
  \frac12 E^2 + t_{44}
  \end{array}
\right),
\end{eqnarray}
where
\begin{eqnarray}
  \begin{array}{lll}
    t_{11}&=&- \frac1{2 \cos \alpha} m_K^2 \\
    t_{12}&=&\frac1{2 \sqrt{3}} (m_K^2-m_\pi^2) \\
    t_{22}&=& -\frac16 (2 m_\pi^2+m_K^2 \cos \alpha) \\
    t_{23}&=&-\frac1{\sqrt{3}} (\frac12 \mu_I + \mu_S) \sin \alpha \\
    t_{34}&=& -i (\frac12 \mu_I + \mu_S) \cos \alpha \\
    t_{44}&=& \frac12  (-m_K^2 \cos \alpha + (\frac12 mu_I + \mu_S )^2 \cos 2 \alpha ),
  \end{array}
\end{eqnarray}
and $\cos \alpha=(m_K/(\frac12\mu_I+\mu_S))^2$, as given in (\ref{kaon}).
One of the modes, the $\tilde K^+$, is massless. The masses
of $\tilde \pi^+, \tilde \pi^-, \tilde K^0, \tilde {\bar K}^0$ are
given by the zeros on the $E$-plane of
\begin{eqnarray}
{\rm det} \left(
\begin{array}{cccc}
\frac12 E^2 + u_1 & E  u_2 & E u_3 & 0 \\
 E u_2^\dagger &
  \frac12 E^2 + u_1  & 0 &  -E u_3 \\
E u_3^\dagger & 0 &
  \frac12 E^2 + u_4 & E u_5 \\
0 & -E u_3^\dagger & E u_5^\dagger&
  \frac12 E^2 + u_4
  \end{array}
\right),
\end{eqnarray}
where
\begin{eqnarray}
  \begin{array}{lll}
    u_1&=& \frac14 ( -2 m_\pi^2 + \mu_I^2 - 2 \mu_I \mu_S +
        2 \mu_I ( \frac12 \mu_I + \mu_S) \cos \alpha ) \\
    u_2&=&  \frac{i}4 (3 \mu_I -
        2 \mu_S + 2 (\frac12 \mu_I + \mu_S) \cos \alpha ) \\
    u_3&=&\frac12 (\frac12 \mu_I + \mu_S) \sin \alpha \\
    u_4&=&\frac18 ( -4 m_K^2 + 2 \mu_I^2 - 4 \mu_I \mu_S -
        \mu_I^2 \cos \alpha + 4 \mu_S^2 \cos \alpha) \\
    u_5&=& \frac{i}4 (-3 \mu_I +
        2 \mu_S + 2 (\frac12 \mu_I + \mu_S) \cos \alpha ),
  \end{array}
\end{eqnarray}
and $\cos \alpha=(m_K/(\frac12\mu_I+\mu_S))^2$, as given in (\ref{kaon}).

\begin{center}
\begin{figure}[ht!]
\vspace{0.5cm}
\hspace*{1cm}
\epsfig{file=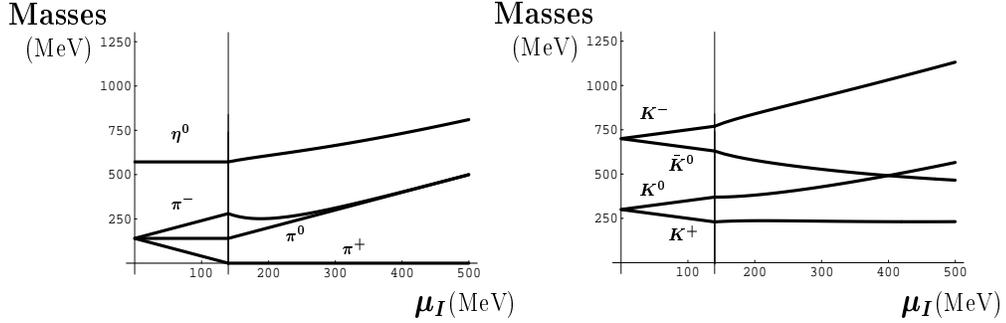, width=14.6cm, height=8.5cm}
\vspace{-4.5cm}
\caption[]{\small The masses of the eight pseudo-Goldstone bosons are
shown here as a function of $\mu_I$ for $\mu_S=200$MeV.
In these two figures, the critical value of
$\mu_I$ at the second order phase transition is depicted by a
solid vertical line.}
\end{figure}
\end{center}

The spectrum is depicted as a function of $\mu_I$ for
$\mu_S=200$MeV in Fig.~4, and for $\mu_S=460$MeV in Fig.~5, the
corresponding condensates and densities are given in Fig.~2, and
in Fig.~3, respectively. The masses are continuous across the
second order phase transition lines, as they should, and they are
in general discontinuous across the first order phase transition
curves, as they might. In each superfluid phase there is one single
massless Goldstone mode. The $SU_V(3) \times U_V(1)$ symmetry 
that leaves the quark-antiquark condensate invariant is further broken
by the new condensates in the superfluid phases:
$U_I(1)$ symmetry is broken by the pion condensate, and $U_S(1)$ 
is broken by the kaon condensate.

\begin{center}
\begin{figure}[ht!]
\vspace{0.5cm}
\hspace*{1cm}
\epsfig{file=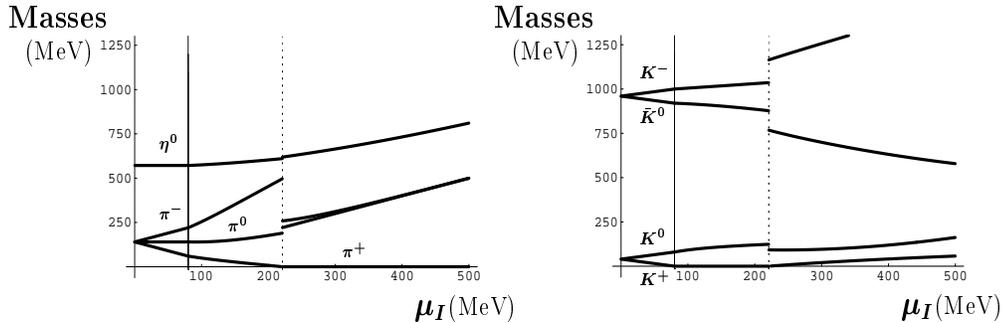, width=14.6cm, height=8.5cm}
\vspace{-4.5cm}
\caption[]{\small The masses of the eight pseudo-Goldstone bosons are
shown here as a function of $\mu_I$ for $\mu_S=460$MeV.
In these two figures, the critical value of
$\mu_I$ at the second order phase transition is depicted by a
solid vertical line, and the critical value of
$\mu_I$ at the first order phase transition is depicted by a
dotted vertical line.}
\end{figure}
\end{center}

\section{Conclusions}
In this article we have explored part of the phase diagram of QCD at
non-zero quark chemical 
potentials and zero temperature. We have used a low-energy effective 
theory based on the symmetries of the QCD partition function, Chiral Perturbation Theory.
The quark chemical potentials have been introduced into the 
effective Lagrangian as a usual vector external source via a flavor gauge symmetry.
Because we only kept
the octet of the Goldstone bosons as relevant degrees of freedom, which is
only justified for small enough chemical potentials, our analysis 
is intrinsicly limited to the domain $|\mu_u|<300 {\rm MeV}$, 
$|\mu_d|<300 {\rm MeV}$, and $|\mu_s|<550 {\rm MeV}$. But this {\it whole}
domain can be reached within Chiral Perturbation Theory.

We have found a non-trivial phase diagram. There are three distinct phases:
a normal phase, a pion superfluid phase, and a kaon superfluid phase. The two
superfluid phases are separated by a first order phase transition curve, whereas
each of them is separated from the normal phase by a second order phase
transition curve. The observables that we have analyzed show a typical behavior 
for such a phase diagram. 
The behavior of the different condensates at the second order phase transitions 
is very similar to what has been found in QCD-like theories \cite{KSTVZ,SSS}. 
The number densities just above the second order phase transition curves can be 
understood from a dilute gas approximation \cite{KSTVZ}.  The excitation spectrum 
in the normal phase is trivial, and 
there is one massless mode in each superfluid phase that is due to  
further spontaneous symmetry breaking by the unusual condensates, 
similar to what as been observed in  QCD-like theories \cite{KST,KSTVZ,SSS}.

Our study extends the analysis of QCD at non-zero isospin chemical potential done by 
Son and Stephanov \cite{SonMisha}. We recover all their results on the 
pion condensate, the isospin number density and the mass spectrum 
for $|\mu_S| < m_K-m_\pi/2=430$MeV,
where kaon condensation can become energetically more favorable, depending 
on $\mu_I$ (cf. Fig.1, 2, 4).

Finally, at low energy and within the effective theory approach we have
used, 
it would be interesting to study matter that is not only made out of quarks and gluons, 
but also electrons and neutrinos. The inclusion of the electro-magnetic interaction 
and of the weak interaction into Chiral Perturbation Theory  has already been 
achieved in the vacuum \cite{LesHouches}, 
and its generalization to non-zero quark chemical potential seems to be rather 
straightforward. The effective theory technique we have used in this article
is therefore very well suited to study more general problems of this kind.

\vskip 2.5cm \noindent {\bf Acknowledgments} \vskip 0.5cm

G. Baym, M. Stephanov and J. Verbaarschot are acknowledged for useful discussions. 
J.B.K. is supported in part by the National Science Foundation, NSF-PHY96-05199. 
D.T. is supported in part by ``Holderbank''-Stiftung.

\end{document}